# Speaker Identification Investigation and Analysis in Unbiased and Biased Emotional Talking Environments

Ismail Shahin

Electrical and Computer Engineering Department

University of Sharjah

P. O. Box 27272

Sharjah, United Arab Emirates

Tel: (971) 6 5050967

Fax: (971) 6 5050877

E-mail: ismail@sharjah.ac.ae

**Abstract**

This work aims at investigating and analyzing speaker identification in each unbiased and biased emotional talking environments based on a classifier called Suprasegmental Hidden Markov Models (SPHMMs). The first talking environment is unbiased towards any emotion, while the second talking environment is biased towards different emotions. Each of these talking environments is made up of six distinct emotions. These emotions are neutral, angry, sad, happy, disgust and fear. The investigation and analysis of this work show that speaker identification performance in the biased talking environment is superior to that in the unbiased talking environment. The obtained results in this work are close to those achieved in subjective assessment by human judges.

**Keywords:** emotional talking environment; hidden Markov models; speaker identification; suprasegmental hidden Markov models.

# 1. Introduction

Speaker recognition is the process of automatically recognizing who is speaking on the basis of individual information embedded in speech signals. Speaker recognition involves two applications: speaker identification and speaker verification (authentication). Speaker identification is the process of finding the identity of the unknown speaker by comparing his/her voice with voices of registered speakers in the database. This type of speaker recognition is a one-to-many comparison. Speaker verification is the process of determining whether the speaker identity is who the person claims to be. Such type of speaker recognition is a one-to-one comparison [1], [2].

Speaker recognition is often classified into closed-set recognition and open-set recognition. The closed-set recognition refers to the cases that the unknown voice must come from a set of known speakers, while the open-set recognition refers to the situations that the unknown voice may come from unregistered speakers. Speaker recognition typically operates in one of two cases: text-dependent (fixed-text) case or text-independent (free-text) case. Text-dependent requires a user to regenerate utterances containing the same text, while text-independent does not require knowledge of the text to be spoken.

Speaker recognition systems perform poorly in emotional talking environments [3], [4], [5], [6], [7]. Speaker recognition by emotion is one of research fields in the human-computer interaction or affective computing [8]. A major incentive comes from the need to develop human machine interface that is more adaptive and responsive to a user's identity. The main task of intelligent human-machine interaction is to enable the machine with the affective computing capability so that the machine can identify the user for many different applications. The applications of speaker identification in emotional talking environments appear in criminal investigations to identify the suspected persons who produced emotional voice and in Text-To-Speech (TTS) communication-aid that can help expressing the emotion of the spoken text [9].

This work is devoted to studying, investigating and analyzing speaker identification in two emotionally unlike and separate speech databases. The first database is unbiased towards any emotional state, while the second database is biased towards different emotional states.

## 2. Motivation and Literature Review

Some earlier studies in the field of speaker recognition focused on speaker recognition in emotional talking environments. Bao *et al*. shed the light on emotion attribute projection for speaker recognition on emotional speech [3]. Wu *et al.* investigated the rules based feature modification for robust speaker recognition with emotional speech [4]. Li *et al.* proposed an approach of speech emotion-state conversion to enhance speaker recognition performance in emotional talking environments [5]. In two of his earlier studies, Shahin focused on using emotions to identify speakers (emotion-dependent speaker identification) [6] and on speaker identification in emotional talking environments [7]. In the first study [6], Shahin achieved an average speaker identification performance of 78.8% (in a closed set with forty speakers and six emotions). In the second study [7], he obtained an average speaker identification performance of 61.4%, 66.4% and 69.1% based, respectively, on hidden Markov models (HMMs), second-order circular hidden Markov models (CHMM2s) and suprasegmental hidden Markov models (SPHMMs) using forty speakers and five emotions.

In the last two decades, a new field called emotion recognition has been introduced and used in our daily life applications such as telecommunications, media and the enhancement of service quality in call centers [10], [11], [12]. In two of his most recent studies, Shahin studied, analyzed and investigated emotion identification when the database is biased towards different emotions based on each of HMMs [13] and SPHMMs [14]. The results of these two studies show that emotion identification performance in a database biased towards emotional states is much better than that in a database unbiased towards any emotion [13], [14].

Our main contribution in this research is focused on studying, investigating and analyzing text-independent and emotion-dependent speaker identification in two distinct and separate emotional speech databases using SPHMMs as a classifier. The first database is unbiased towards any emotion, while the second database is biased towards different emotions. Six basic emotions have been used to study speaker identification in each emotional talking environment. These emotions are neutral, angry, sad, happy, disgust and fear. The current study is different from our two previous studies in [13] and [14]. In this work, we focus on studying, investigating and analyzing speaker identification in completely two different and separate emotional talking environments, while the two studies in [13] and [14] focused on studying, investigating and analyzing emotion identification in completely two unlike and separate databases based on HMMs and SPHMMs, respectively. To the best of our knowledge, this is the first attempt to study and investigate speaker identification in an environment biased towards different emotional states.

The remainder of this paper is organized as follows. The next section overviews the basics of SPHMMs. Section 4 describes the two emotional speech databases used in this work and the adopted features. The algorithm that has been used in identifying speakers and the experiments are discussed in Section 5. Section 6 demonstrates the results obtained in this work and their discussion. Concluding remarks are given in Section 7.

## 3. Basics of Suprasegmental Hidden Markov Models

SPHMMs have been developed, used and evaluated by Shahin in the fields of speaker recognition [7], [15] and emotion recognition [16]. SPHMMs proved to be superior models over HMMs for speaker recognition in each of emotional [7] and shouted talking environments [15]. SPHMMs possess the ability to condense several states of HMMs into what is designated suprasegmental state. Suprasegmental state can look at the observation sequence through a larger window. Such a state allows observations at rates appropriate for the situation of modeling emotional speech signals. For example, prosodic information can not be perceived at a rate that is used for acoustic modeling. Within HMMs, combining prosodic and acoustic information can be achieved as given by the following formula [17],

$$log\ P\left(\lambda^v, \Psi^v \mid O\right) = (1-\alpha)\cdot log\ P\left(\lambda^v \mid O\right) + \alpha \cdot log\ P\left(\Psi^v \mid O\right) \qquad (1)$$

where $\alpha$ is a weighting factor. When

$$\begin{cases} 0.5 > \alpha > 0 & \text{biased towards acoustic model} \\ 1 > \alpha > 0.5 & \text{biased towards prosodic model} \\ \alpha = 0 & \text{biased completely towards acoustic model and} \\ & \text{no effect of prosodic model} \\ \alpha = 0.5 & \text{not biased towards any model} \\ \alpha = 1 & \text{biased completely towards prosodic model and} \\ & \text{no impact of acoustic model} \end{cases} \qquad (2)$$

$\lambda^v$ is the acoustic model of the $v^{th}$ speaker, $\Psi^v$ is the suprasegmental model of the $v^{th}$ speaker, $O$ is the observation vector or sequence of an utterance, $P\left(\lambda^v \mid O\right)$ and $P\left(\Psi^v \mid O\right)$ can be computed using Bayes theorem as given in Eqs. (3) and (4), respectively [18],

$$P\left(\lambda^v \mid O\right) = \frac{P\left(O \mid \lambda^v\right) P_0\left(\lambda^v\right)}{P(O)} \qquad (3)$$

$$P\left(\Psi^v \mid O\right) = \frac{P\left(O \mid \Psi^v\right) P_0\left(\Psi^v\right)}{P(O)} \qquad (4)$$

where $P_0(\lambda^v)$ and $P_0(\Psi^v)$ are, respectively, the priori distribution of the acoustic model and the suprasegmental model. More information about suprasegmental hidden Markov models can be obtained from [7], [15], [16].

## 4. Speech Databases and Extraction of Features

### 4.1. Speech Databases

In this work, speaker identification has been studied, investigated and analyzed in two distinct and separate speech databases. The first database is unbiased towards any emotional state, while the second one is biased towards different emotional states.

The first database was collected from fifty (twenty five male and twenty five female) untrained healthy adult native speakers of American English. In this database, each speaker was asked to utter five sentences where each sentence was uttered fifteen times (nine times in the training session and the remaining in another separate session called the test or identification session) under each of the neutral, angry, sad, happy, disgust and fear emotions. In this database, speakers uttered the desired sentences naturally. These speakers were allowed to hear some recorded sentences before uttering the required database. The speakers were not allowed to practice generating such sentences under any emotion. In such a database, the five sentences were unbiased towards any emotion. These sentences are:

1) *He works five days a week.*
2) *The sun is shining.*
3) *The weather is fair.*
4) *Assistant professors are looking for promotion.*
5) *Electrical and computer engineering department.*

The above five sentences were unbiased towards any emotion which means that the content of these sentences was not biased towards any emotion (*i.e.* there was no correlation between any sentence and any emotion). For example, the content of the last sentence "electrical and computer engineering department" can not be biased towards any emotion.

The second database was gathered from the same fifty speakers. The fifty speakers were asked to generate different five sentences under each of the neutral, angry, sad, happy, disgust and fear emotions where each sentence was uttered fifteen times (nine times in the training session and the rest in the test session). Each different five sentences were biased towards their corresponding emotion (*i.e.* there was a correlation between each five sentences and their corresponding emotion).

The five sentences under neutral state are the same as the previous five sentences. The five sentences biased towards angry emotion are:

1) *Do not override my decision.*
2) *Stop talking during the lecture.*
3) *You broke my laptop.*
4) *His behavior made me angry.*
5) *Your decision was wrong.*

The last five sentences were biased towards angry emotion. The content of every sentence is biased towards this emotion. For example, the first sentence "do not override my decision" is not expected for someone to hear it from speakers speaking in neutral, sad, happy, disgust, or fear emotion. This sentence can only be heard when speakers are in an anger mode.

The five sentences biased towards sad emotion are:

1) *He passed away few hours ago.*
2) *My son failed in the TOEFL exam.*
3) *Sorry to hear the bad news.*
4) *My friend was fired from his job.*
5) *The decision was unfair.*

The five sentences biased towards happy emotion are:

1) *I was happy when I went back home.*
2) *My son passed the TOEFL exam successfully.*
3) *Good to hear the good news.*
4) *It is my pleasure to meet you.*
5) *The decision was fair.*

The five sentences biased towards disgust emotion are:

1) *He vomited in my car.*
2) *That was too terrible.*
3) *His case shocked me.*
4) *The refrigerator smells very bad.*
5) *Your talk was disgusting.*

The five sentences biased towards fear emotion are:

1) *The story scared me.*
2) *I am afraid from failure.*
3) *The results are alarming.*
4) *I am worry about my promotion.*
5) *That was a nightmare for me.*

Our speech databases were recorded in a clean environment that was not affected by a background noise. The two databases were separately captured by a speech acquisition board using a 16-bit linear coding A/D converter and sampled at a sampling rate of 16 kHz. These databases were 16-bit per sample linear data. The speech signals were applied every 5 ms to a 30 ms Hamming window. The two speech databases used in this work were "closed set".

## 4.2. Extraction of Features

In the present work, the features that have been adopted to represent the phonetic content of speech signals are called the short time Log Frequency Power Coefficients (LFPCs). LFPCs have proven to be superior features over each of the Linear Prediction Cepstral Coefficients (LPCCs) and the Mel-Frequency Cepstral Coefficients (MFCCs) in emotional talking environments [14], [19].

The windowed speech signal, after applying the Hamming window, was transformed into the frequency domain using the Discrete Fourier Transform (DFT) algorithm. The spectral components were separated into 16 bands. The DFT responses of the 16 filters were nothing but shifting and frequency warping versions of a rectangular window $W_m(k)$ which is given as [19],

$$W_m(k) = \begin{cases} 1 & h_m \geq k \geq l_m, \quad m = 1, 2, \ldots, 16 \\ 0 & \text{otherwise} \end{cases} \quad (5)$$

where $k$ is the DFT domain index, $l_m$ and $h_m$ are, respectively, the lower and upper edges of the $m$th filter bank which is given by,

$$S_t(m) = \sum_{k=f_m-(b_m/2)}^{k=f_m+(b_m/2)} \left[X_t(k)\, W_m(k)\right]^2, \quad m = 1, 2, \ldots, 16 \quad (6)$$

where $X_t(k)$ is the $k$th spectral component of the windowed signal, $t$ is the frame number, $S_t(m)$ is the output of the $m$th filter bank, $f_m$ and $b_m$ are the center frequency of the $m$th sub-band and the bandwidth of the $m$th sub-band, respectively. For each frame, 16 LFPCs were obtained.

In the current work, an ergodic or fully connected SPHMM structure that has been obtained from an ergodic HMM structure becomes more appropriate than a left-to-right structure because the emotional cues contained in an emotional utterance cannot be assumed as specific sequential events in the signal. It is well known that every state in the ergodic structure can be reached in a single step from every other state [18]. In this work, the number of suprasegmental states was three, while the number of conventional states was nine (each suprasegmental state was comprised of three conventional states). The number of mixture components, $M$, was ten per state with a continuous mixture observation density was selected for the recognizer. Fig. 1 shows our adopted three-state ergodic SPHMM structure that was derived from a nine-state ergodic HMM structure.

## 5. Speaker Identification Algorithm Based on SPHMMs and the Experiments

### 5.1. Unbiased Emotional Talking Environment

In this talking environment, the training session based on SPHMMs is similar to the training session based on HMMs. In the training session of SPHMMs, suprasegmental models were trained on top of acoustic models. In this session, the $v$th SPHMM speaker model was derived using all the five sentences uttered under each emotion with a repetition of nine utterances / sentence. The total number of utterances used to construct the $v$th speaker model in such a session was 270 (5 sentences X 6 emotions X 9 utterances / sentence).

In the evaluation (identification) session of such talking environment (completely separate from the training session), each one of the fifty speakers used six of the fifteen utterances per sentence (text-independent) under each emotion (emotion-dependent). The total number of utterances used in this session was 9000 (50 speakers X 6 emotions X 5 sentences X 6 utterances / sentence). The probability of generating every utterance was computed based on SPHMMs (there were 50 probabilities). The model with the highest probability was chosen as the output of speaker identification as given in the following formula,

$$V^* = \arg\max_{50 \geq v \geq 1} \left\{ P\left(O | \Psi^v\right) \right\} \quad (7)$$

where $V^*$ is the index of the identified speaker, $O$ is the observation vector or sequence that belongs to the unknown speaker, $P(O|\Psi^v)$ is the probability of the observation sequence $O$ given the $v$th SPHMM speaker model $\Psi^v$. This session can be summarized in the block diagram given in Fig. 2.

**5.2. Biased Emotional Talking Environment**

In the training session of this talking environment, there are six separate training sessions each based on SPHMMs. The first training session belongs to the five sentences under neutral state. This training session is the same as that in the unbiased emotional talking environment.

The second training session belongs to the five sentences biased towards angry emotion. In this session, the $v$th SPHMM speaker model was built using all the five sentences biased towards angry emotion in addition to all the five sentences uttered under each unbiased emotion excluding the angry emotion. The training data in this phase was composed of 270 utterances ((5 sentences biased towards angry emotion X 9 utterances / sentence) + (5 unbiased sentences X 5 emotions X 9 utterances / sentence)).

The third training session corresponds to the five sentences biased towards sad emotion. The $v$th SPHMM speaker model in this session was constructed by exploiting all the five sentences biased towards sad emotion plus all the five sentences generated under each unbiased emotion except the sad emotion. The training data in such a session was made up of 270 utterances.

The fourth training session belongs to the five sentences biased towards happy emotion. In such a stage, the $v$th SPHMM speaker model was obtained by employing all the five sentences biased towards happy emotion along with all the five sentences generated under each unbiased emotion apart from the happy emotion. The training data in this session was comprised of 270 utterances.

The fifth training stage is associated with the five sentences biased towards disgust emotion. In this stage, the $v$th SPHMM speaker model was achieved using all the five sentences biased towards disgust emotion in conjunction with all the five sentences produced under each unbiased emotion aside from the disgust emotion. The training data in such a session consisted of 270 utterances.

Finally, the sixth training phase is linked to the five sentences biased towards fear emotion. The $v$th SPHMM speaker model in such a phase was attained using all the five sentences biased towards fear emotion in addition to all the five sentences produced under each unbiased emotion excluding the fear emotion. The training data in such a phase was composed of 270 utterances.

In the evaluation session of the biased talking environment, there are six separate evaluation sessions each based on SPHMMs. The first evaluation session belongs to the first training session. The second evaluation session belongs to the second training session, *etc*…. In each evaluation session, each one of the fifty speakers used six of the fifteen utterances per sentence (text-independent) under each emotion. The total number of utterances used in each session was 9000 (50 speakers X 6 emotions X 5 sentences X 6 utterances / sentence). In each separate session, the probability of generating every utterance was computed based on SPHMMs. The model with the highest probability was selected as the output of speaker identification.

## 6. Results and Discussion

### 6.1. Speaker Identification in the Unbiased Emotional Talking Environment

Speaker identification performance based on SPHMMs in the unbiased talking environment is summarized in Table 1. The average speaker identification performance of this table is 73.08%. The table shows that the highest speaker identification performance happens when speakers speak in a neutral state (86.5%), while the least speaker identification performance occurs when speakers speak in an angry emotion (64.5%). The results attained in this work are consistent with those obtained by Shahin in one of his studies [7]. He achieved, based on the same classifier and using forty speakers uttering eight different sentences under five different unbiased emotions, an average speaker identification performance of 69.10%.

To make a comparison between HMMs and SPHMMs as classifiers in identifying the unknown speaker in the unbiased emotional talking environment, SPHMMs have been replaced with HMMs. Table 2 shows speaker identification performance based on HMMs in such an environment. This table gives average speaker identification performance of 66.33%. It is evident from the results obtained in this table and those obtained in Table 1 that speaker identification performance based on SPHMMs in such emotional talking environment is better than that based on HMMs evaluated in the same environment. This may be attributed to the following reasons:

1. HMMs are not convenient enough as a classifier for speaker identification in emotional talking environment.

2. SPHMMs have the ability to integrate obsevations from emotional modality because such models allow for observations at an appropriate rate for emotional features.

The results achieved in the present work based on HMMs agree with those reported by Shahin using the same models. Shahin obtained an average speaker identification performance of 61.40% using forty speakers uttering eight distinct sentences under five different unbiased emotions [7].

To investigate whether speaker identification performance differences (speaker identification performance based on SPHMMs and that based on HMMs) in the unbiased emotional talking environment are real or simply due to statistical fluctuations, a statistical significance test has been carried out. The statistical significance test has been performed based on the Student *t* Distribution test as given by the following formula,

$$t_{SPHMMs, HMMs} = \frac{\bar{x}_{SPHMMs} - \bar{x}_{HMMs}}{SD_{pooled}} \quad (8)$$

where $\bar{x}_{SPHMMs}$ is the mean of the first sample (based on SPHMMs) of size *n*, $\bar{x}_{HMMs}$ is the mean of the second sample (based on HMMs) of the same size, $SD_{pooled}$ is the pooled standard deviation of the two samples given as,

$$SD_{pooled} = \sqrt{\frac{SD_{SPHMMs}^2 + SD_{HMMs}^2}{n}} \quad (9)$$

where $SD_{SPHMMs}$ is the standard deviation of the first sample (based on SPHMMs) of size *n*, $SD_{HMMs}$ is the standard deviation of the second sample (based on HMMs) of the same size.

Based on Table 1 and Table 2, $\bar{x}_{SPHMMs\_unb} = 73.08\%$, $SD_{SPHMMs\_unb} = 7.36$, $\bar{x}_{HMMs\_unb} = 66.33\%$, $SD_{HMMs\_unb} = 8.25$. Based on these values, the calculated *t* value is $t_{SPHMMs\_unb,\ HMMs\_unb}$ = 8.191. This calculated *t* value is much greater than the tabulated critical value at *0.05* significant level $t_{0.05}$ = 1.645. Therefore, it is evident from Table 1 and Table 2 that SPHMMs are superior to HMMs for speaker identification in the unbiased emotional talking environment.

**6.2. Speaker Identification in the Biased Emotional Talking Environment**

Table 3 demonstrates speaker identification performance when the talking environment is biased towards angry emotion. This table yields an average speaker identification performance of 76.75%. It is apparent from Table 1 and Table 3 that angry speaker identification performance has been significantly improved by 19.38% when speakers speak in an environment biased towards angry emotion compared to that in an unbiased environment. The two tables also show that speaker identification performance has been improved under all emotions when speakers generate their voices in an environment biased towards angry emotion compared to that in the unbiased talking environment.

Table 4 shows speaker identification performance when the talking environment is biased towards sad emotion. This table gives an average speaker identification performance of 75.58%. It is evident from Table 1 and Table 4 that sad speaker identification performance has been remarkably enhanced when speakers produce their voices in an environment biased towards sad emotion compared to that in an unbiased talking environment. The enhancement rate of speaker identification performance is 16.67%.

Speaker identification performance when the talking environment is biased towards happy emotion is given in Table 5. Based on this table, the average speaker identification performance is 76.08%. Comparing this table with Table 1, happy speaker identification performance when the talking environment is biased towards such an emotion has been appreciably improved by 15.07% compared to that when the talking environment is unbiased towards any emotion.

Average speaker identification performance when the talking environment is biased towards disgust emotion is 77.00% as calculated from Table 6. This table shows that disgust speaker identification performance when the talking environment is biased towards such an emotion has been considerably improved by 17.24% compared to that when the talking environment is unbiased towards any emotion.

Table 7 yields speaker identification performance when the talking environment is biased towards fear emotion. This table gives an average speaker identification performance of 76.33%. It is evident that fear speaker identification performance has been drastically enhanced by 15.75% when the talking environment is biased towards fear emotion compared to that when the talking environment is unbiased towards any emotion.

Three more experiments have been separately conducted to evaluate the results obtained in this work. The three experiments are:
1) An informal subjective evaluation of speaker identification performance in each of the unbiased and biased emotional talking environments has been carried out by ten listeners (human judges) in order to assess speaker identification performance in each talking environment. There are one assessment in the unbiased talking environment and five separate assessments in the biased talking environment. A total of 1500 utterances (fifty speakers X six emotions X five sentences) per assessment have been used. During the six separate assessments, each listener was asked to identify the unknown speaker. Based on the listening tests, the average speaker identification performance when the talking environment is unbiased towards any emotion is 75.25%, while the average speaker identification performances when the talking environment is biased towards angry, sad, happy, disgust and fear emotions are 76.44%, 75.21%, 76.09%, 77.78% and 75.89%, respectively.

The quality of emotions delivered by subjects has been tested and compared based on each of SPHMMs and listening tests by measuring kappa statistic as calculated in Fig. 3. Kappa coefficient is a statistical measure of the actual agreement as compared to the agreement that may happen just by chance [20]. Large values of kappa support the hypothesis that there is a strong relation between two parts (close results), while small values indicate that the observed relation is not more than what is expected by chance. Based on this figure, it is evident that for speaker identification performance based on SPHMMs, there is a moderate agreement for each of the six biased talking environments since kappa values range from 0.302 to 0.408. Therefore, the agreement is due to an actual relation and does not come by chance. The same can be fairly said about the listening tests for each of the six biased talking environments (kappa values range from 0.348 to 0.400). In addition, it can be seen from this figure that the two sets of kappa values per each biased talking environment (the one based on SPHMMs and the one based on the listening test) are relatively close.

2) A statistical cross-validation technique has been performed to estimate the standard deviation of the recognition rates when the talking environment is biased towards each emotion. Cross-validation technique has been implemented per each biased talking environment as follows. The entire collected database (22500 utterances) is partitioned at random into five subsets. Each subset is composed of 4500 utterances. Based on these five subsets per each biased emotional talking environment and using SPHMMs, the standard deviation per biased talking environment is calculated. The standard deviation values are summarized in Fig. 4.

Based on this figure, cross-validation technique shows that the calculated values of standard deviation are very low. Therefore, it is apparent that speaker identification performance based on SPHMMs when the talking environment is biased towards different emotional states (using the five subsets) is very close to that using the entire database (very slight fluctuations).

3) SPHMMs have been replaced as a classifier by HMMs to make a comparison between SPHMMs and HMMs in identifying the unknown speaker in each biased emotional talking environment. Using HMMs as a classifier, the average speaker identification performance when the talking environment is biased towards angry, sad, happy, disgust, and fear emotions is 68.02%, 69.51%, 67.95%, 69.89% and 69.57%, respectively. Table 8 summarizes the calculated $t_{SPHMMs\_biased,\ HMMs\_biased}$ values when the talking environment is biased towards every emotion. This table shows that each calculated $t$ value is much greater than the tabulated critical value $t_{0.05} = 1.645$. It is clear from this table that SPHMMs are superior to HMMs in the biased emotional talking environment.

## 7. Concluding Remarks

In this work, speaker identification has been studied, investigated and analyzed in two separate and unlike emotional talking environments. Some concluding remarks can be drawn from this work. Firstly, speaker identification performance has been significantly improved in the biased emotional talking environment compared to that in the unbiased emotional talking environment. This may be attributed to the fact that speakers usually use certain words and phrases more frequently in expressing their emotions since they have learned the connection between certain words and their corresponding emotions. Therefore, it is easier to identify the unknown speaker in the biased emotional talking environment than in the unbiased talking environment. Secondly, speaker identification performance in the biased emotional talking environment has been enhanced under all emotions. Thirdly, SPHMMs outperform HMMs for speaker identification systems in each unbiased and biased emotional talking environments.

Speaker identification performance in the biased emotional talking environment is limited. This limitation is accredited to the choice of the biased database. The choice of the biased database is not ideal since some sentences are biased towards more than one emotion. Another limitation is that the speech database used in this work is a local one. To the best of our knowledge, there is no global and well-known speech database biased towards different emotional states.

For future work, we plan to thoroughly study and investigate the factors that affect the performance of speaker identification in the biased emotional talking environment. We also plan to propose and implement new classifiers to identify speakers in a such talking environment. I addition, we intend to determine the features that best model the phonetic content of speech signals in the biased talking environment.

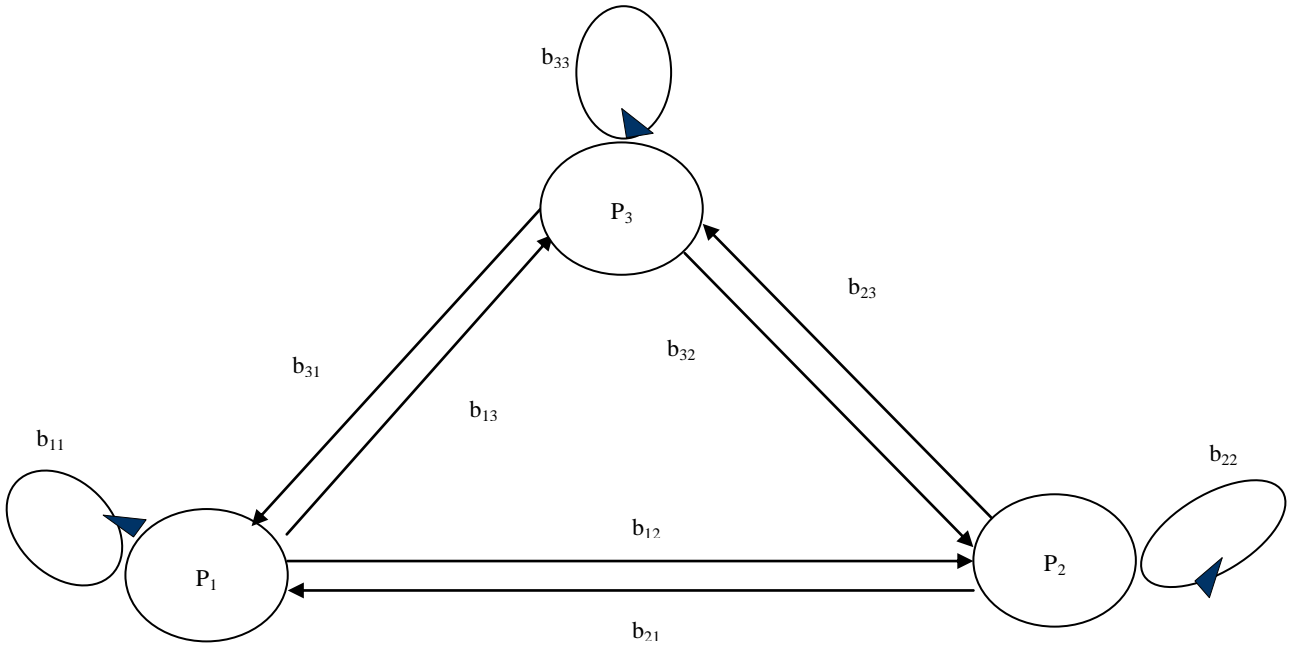

**Figure 1.** Three-state ergodic SPHMM structure derived from a nine-state ergodic HMM structure.

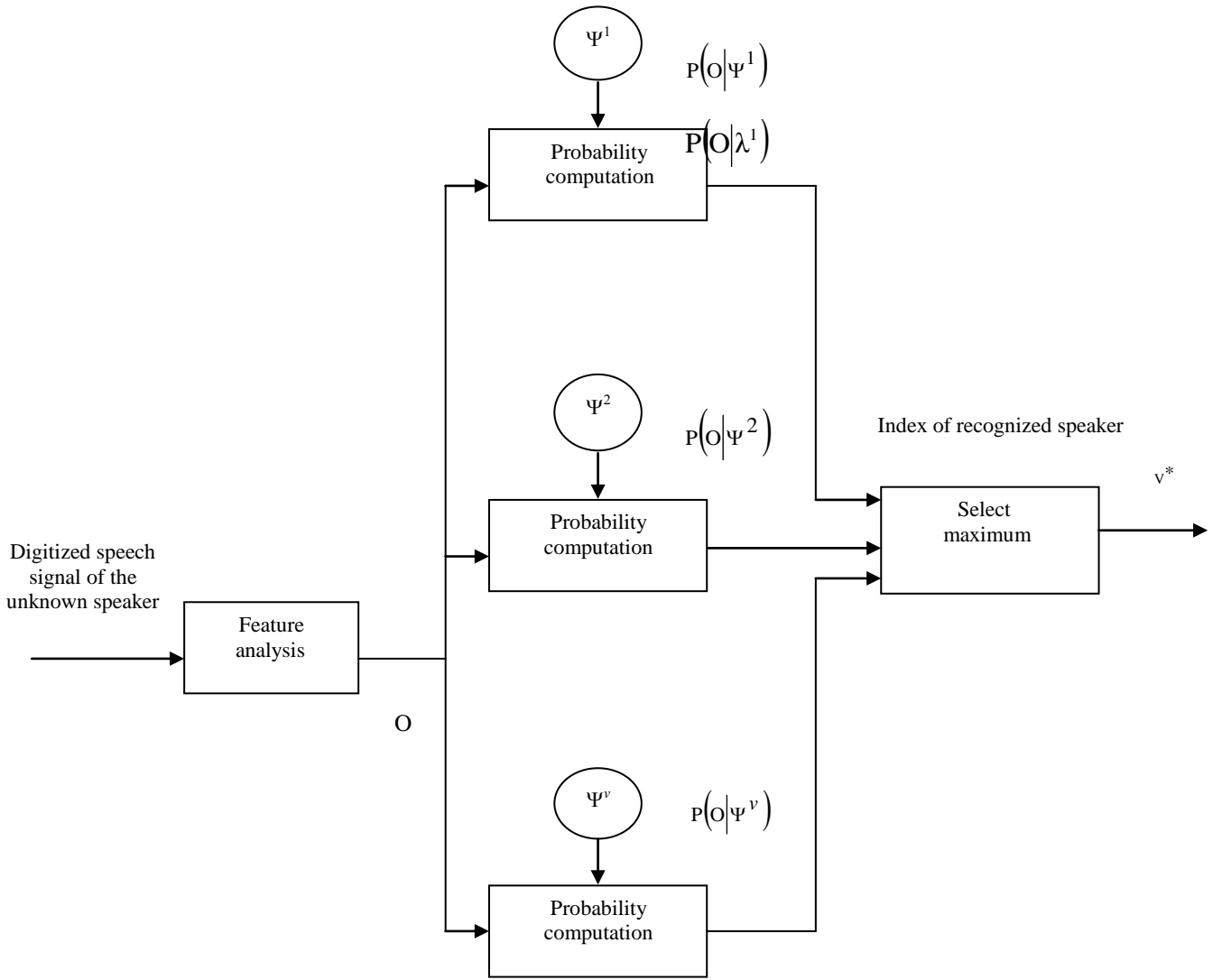

**Figure 2.** Block diagram of speaker identification based on SPHMMs.

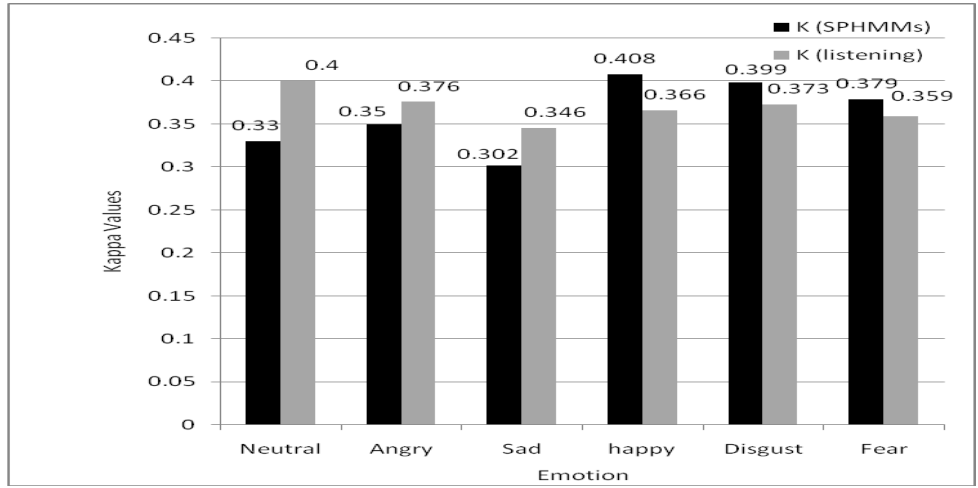

**Figure 3.** Kappa statistic based on each of SPHMMs and listening tests in a talking environment biased towards each of neutral, angry, sad, happy, disgust and fear emotions.

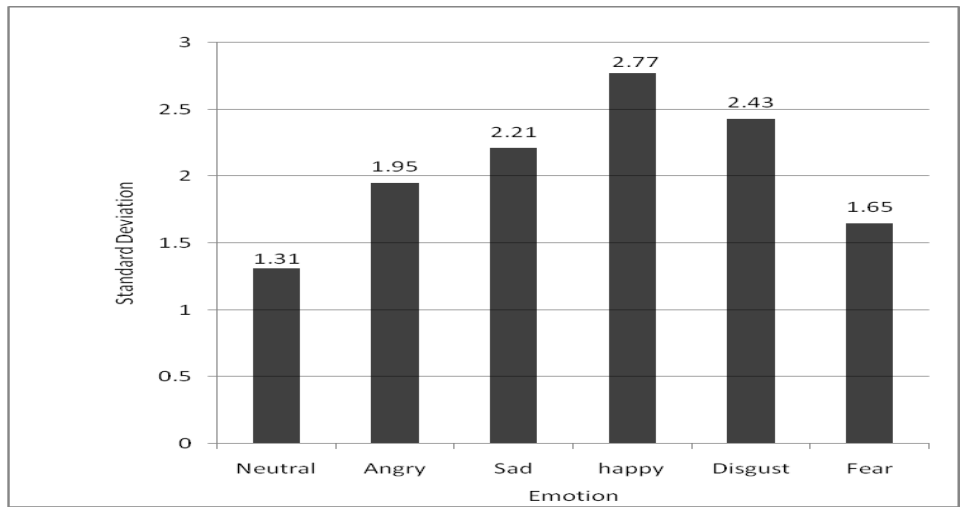

**Figure 4.** Statistical cross-validation technique based on SPHMMs per each biased emotional talking environment.

Table 1
Speaker identification performance based on SPHMMs in the unbiased emotional talking environment

| Emotion | Males (%) | Females (%) | Average (%) |
|---|---|---|---|
| Neutral | 86 | 87 | 86.5 |
| Angry | 64 | 65 | 64.5 |
| Sad | 68 | 70 | 69 |
| Happy | 72 | 74 | 73 |
| Disgust | 73 | 72 | 72.5 |
| Fear | 72 | 74 | 73 |

Table 2
Speaker identification performance based on HMMs in the unbiased emotional talking environment

| Emotion | Males (%) | Females (%) | Average (%) |
|---|---|---|---|
| Neutral | 81 | 82 | 81.5 |
| Angry | 57 | 58 | 57.5 |
| Sad | 61 | 61 | 61 |
| Happy | 65 | 66 | 65.5 |
| Disgust | 67 | 68 | 67.5 |
| Fear | 65 | 65 | 65 |

Table 3
Speaker identification performance based on SPHMMs when the talking environment is biased towards angry emotion

| Emotion | Males (%) | Females (%) | Average (%) |
|---|---|---|---|
| Neutral | 87 | 88 | 87.5 |
| Angry | 76 | 78 | 77 |
| Sad | 70 | 69 | 69.5 |
| Happy | 75 | 76 | 75.5 |
| Disgust | 76 | 74 | 75 |
| Fear | 76 | 76 | 76 |

Table 4
Speaker identification performance based on SPHMMs when the talking environment is biased towards sad emotion

| Emotion | Males (%) | Females (%) | Average (%) |
|---|---|---|---|
| Neutral | 87 | 89 | 88 |
| Angry | 64 | 64 | 64 |
| Sad | 80 | 81 | 80.5 |
| Happy | 73 | 72 | 72.5 |
| Disgust | 74 | 76 | 75 |
| Fear | 74 | 73 | 73.5 |

Table 5
Speaker identification performance based on SPHMMs when the talking environment is biased towards happy emotion

| Emotion | Males (%) | Females (%) | Average (%) |
|---|---|---|---|
| Neutral | 88 | 89 | 88.5 |
| Angry | 64 | 65 | 64.5 |
| Sad | 70 | 69 | 69.5 |
| Happy | 84 | 84 | 84 |
| Disgust | 75 | 73 | 74 |
| Fear | 76 | 76 | 76 |

Table 6
Speaker identification performance based on SPHMMs when the talking environment is biased towards disgust emotion

| Emotion | Males (%) | Females (%) | Average (%) |
|---|---|---|---|
| Neutral | 89 | 89 | 89 |
| Angry | 68 | 68 | 68 |
| Sad | 70 | 72 | 71 |
| Happy | 74 | 73 | 73.5 |
| Disgust | 85 | 85 | 85 |
| Fear | 75 | 76 | 75.5 |

Table 7
Speaker identification performance based on SPHMMs when the talking environment is biased towards fear emotion

| Emotion | Males (%) | Females (%) | Average (%) |
|---|---|---|---|
| Neutral | 88 | 88 | 88 |
| Angry | 66 | 66 | 66 |
| Sad | 71 | 71 | 71 |
| Happy | 74 | 74 | 74 |
| Disgust | 75 | 74 | 74.5 |
| Fear | 84 | 85 | 84.5 |

Table 8
Calculated $t$ values when the talking environment is biased towards each emotion

| Talking environment biased towards: | Calculated $t$ value $t_{\text{SPHMMs, HMMs}}$ |
|---|---|
| Angry | 8.312 |
| Sad | 8.911 |
| Happy | 8.433 |
| Disgust | 8.001 |
| Fear | 8.453 |